\documentclass[showkeys,superscriptaddress,floatfix,prd,10pt,aps]{revtex4-2}
\usepackage{graphicx,epstopdf}
\usepackage[caption=false]{subfig}
\usepackage{appendix}
\usepackage[T1]{fontenc}
\usepackage{lmodern}
\usepackage[dvipsnames,x11names]{xcolor}
\usepackage[colorlinks=true,linkcolor=NavyBlue,citecolor=ForestGreen,urlcolor=NavyBlue]{hyperref}
\usepackage[sort&compress]{natbib}
\usepackage{lipsum}
\usepackage{morefloats}
\usepackage[pdf]{pstricks}
\usepackage{amsmath}
\usepackage{amssymb}
\usepackage{amsfonts}
\usepackage{rotating}
\usepackage{cancel}
\usepackage{mathtools}
\usepackage{bbm}
\usepackage{dsfont}
\usepackage{bbold}
\usepackage{multirow}
\usepackage{ulem}
\usepackage{physics}
\usepackage{orcidlink}
\usepackage{colortbl}

\def\pslash{p\!\!\!\slash }
\def\p_1slash{p1\!\!\!\slash }
\def\p_2slash{p2\!\!\!\slash }
\def\qslash{q\!\!\!\slash }
\def\xslash{x\!\!\!\slash }
\def\eslash{\varepsilon\!\!\!\slash }

\def\vel{\left|}
\def\ver{\right|}

\begin{document}

\title{Insight into the nature of the $P_{c}(4457)$ and related pentaquarks}

\author{Ula\c{s}~\"{O}zdem\orcidlink{0000-0002-1907-2894}}%
\email[]{ulasozdem@aydin.edu.tr }
\affiliation{ Health Services Vocational School of Higher Education, Istanbul Aydin University, Sefakoy-Kucukcekmece, 34295 Istanbul, T\"{u}rkiye}

\begin{abstract}
We systematically study the electromagnetic properties of pentaquark states from different perspectives to better understand their nature, internal structure, and quantum numbers, determine their hadronization processes, and shed light on their true nature.  The present study examines the magnetic moments of the $P_{c}(4457)$ and related hidden-charm pentaquark states with and without strangeness ($[d d][u c] \bar c$, $[u u][s c] \bar c$, $[dd ][s c] \bar c$, $[s s][u c] \bar c$ and $[s s][d c] \bar c$), employing a comprehensive analysis that encompasses both the compact pentaquark configuration and $J^P = \frac{3}{2}^-$ quantum numbers. The present study compares the results regarding the magnetic moment of the $P_{c}(4457)$ pentaquark state with those reported in the existing literature. The numerical results obtained in this study, when considered alongside existing literature, indicate that the magnetic moments of hidden-charm pentaquark states may offer insights into their underlying structures, which in turn can inform the distinction between their spin-parity quantum numbers. It seems that for the future experimental search of the family of hidden-charm pentaquark states, studying the electromagnetic properties of the hidden-charm pentaquark states can provide valuable information.
\end{abstract}

\maketitle

\section{Introduction}\label{motivation}

The investigation of exotic states, such as tetraquarks, hybrids, glueballs,  and pentaquarks, has become a prominent area of focus in hadron physics following the proposal of the quark model. Given that neither the quark model nor QCD prohibited their existence, these states attracted attention from the outset and were subjected to extensive investigation over an extended period. Ultimately, expectations were fulfilled by announcing the first discovery of such states in 2003, namely a tetraquark state, X(3872), by the Belle Collaboration \cite{Belle:2003nnu}.  Subsequently, the number of observed exotic states increased and their diversity expanded, following the findings yielded by the aforementioned experimental discovery.  A thorough investigation of these exotic states may yield substantial insights into the fundamental processes underlying the dynamics of strong interactions at low energies.  In 2015, a novel member of the exotic states, namely the pentaquark state comprising five valence quarks, was reported to have been discovered by the LHCb Collaboration. The two states, designated as $P_c(4380)^+$ and $P_c(4450)^+$, were confirmed through observation in the $J/\psi+p$ decay channel \cite{LHCb:2015yax}. In 2019, the analyses with a larger data sample yielded further insights. It was revealed that the previously reported  $P_c(4450)^+$ state had split into $P_c(4440)^+$ and $P_c(4457)^+$ states, and another pick, $P_c(4312)^+$, had also come into sight \cite{LHCb:2019kea}.
It should be noted that the pentaquark $P_c(4380)^+$ reported in the previous analysis remains unresolved, neither confirmed nor refuted, in the subsequent analysis. The six-dimensional amplitude analysis presented in Ref. \cite{LHCb:2015yax}, which initially provided evidence for the $P_c(4380)^+$ state, is now considered obsolete, as it included only a single $P_c(4450)^+$ state and did not account for the $P_c(4312)^+$ state. Consequently, the results presented in this Letter weaken the previously reported evidence for the $P_c(4380)^+$ state; however, they do not contradict its existence, since the current one-dimensional analysis lacks sensitivity to broad $P_c^+$ states.  A future six-dimensional amplitude analysis of \( \Lambda_b \to J/\psi p K^- \) decays, incorporating the \( P_c(4440)^+ \), \( P_c(4457)^+ \), and \( P_c(4312)^+ \) states, will be essential to assess whether there is continuing evidence for the \( P_c(4380)^+ \) state or any other broad \( P_c^+ \) states. It is worth mentioning that Ref.~\cite{Du:2019pij} demonstrated evidence for a narrow \( \Sigma_c^* \bar{D} \) bound state in the $J/\psi p$ invariant mass distribution data, which they refer to as \( P_c(4380)^+ \) with spin-parity $J^P=\frac{3}{2}^-$, distinct from the broad structure reported by LHCb in 2015. 
 In 2020, the LHCb Collaboration announced a pentaquark state, $P_{cs}(4459)^0$, in the invariant mass spectrum of $J/\psi\Lambda$ in the $\Xi_b^0 \rightarrow J/\psi\,\Lambda\,K^-$ decay \cite{LHCb:2020jpq}. The measured mass and width are $4458.8 \pm 2.7 ^{+4.7}_{-1.1}$ MeV and $17.3 \pm 6.5^{+8.0}_{-5.7}$ MeV respectively.   In 2022, the LHCb collaboration observed a new structure $P_{cs}(4338)^0$ in the $J/\psi\Lambda$ mass distribution in the $B^- \rightarrow J/\psi\Lambda^- p $ decays \cite{LHCb:2022ogu}. The masses, widths, minimal valence quark contents, and observed channels for these states have been listed in Table \ref{pentaquarks}.

%
\begin{table}[htp]
\caption{Hidden-charm pentaquark states reported by the LHCb
Collaboration.}\label{pentaquarks}
\begin{tabular}{l|c|c|c|c}
\toprule
State  & Mass (MeV) & Width (MeV) & Content &  Observed channels\\
\toprule
$P_c(4380)^+$ \cite{LHCb:2015yax}  &            $4380\pm8\pm29$          &       $215\pm18\pm86$           & $uudc\bar{c}$ & $\Lambda_b^0 \to J/\psi pK^-$\\
$P_c(4312)^+$ \cite{LHCb:2019kea}       & ~~$4311.9\pm0.7^{~+6.8}_{~-0.6}$~~  &  $9.8\pm2.7^{~+3.7}_{~-4.5}$    & $uudc\bar{c}$ & $\Lambda_b^0 \to J/\psi pK^-$ \\
$P_c(4440)^+$ \cite{LHCb:2019kea}      &    $4440.3\pm1.3^{~+4.1}_{~-4.7}$   &  $20.6\pm4.9^{~+8.7}_{~-10.1}$  & $uudc\bar{c}$& $\Lambda_b^0 \to J/\psi pK^-$ \\
$P_c(4457)^+$  \cite{LHCb:2019kea}    &    $4457.3\pm0.6^{~+4.1}_{~-1.7}$   &  $6.4\pm2.0^{~+5.7}_{~-1.9}$    & $uudc\bar{c}$ & $\Lambda_b^0 \to J/\psi pK^-$ \\
$P_{cs}(4459)^0$ \cite{LHCb:2020jpq}   &    $4458.8\pm2.9^{~+4.7}_{~-1.1}$   &  $17.3\pm6.5^{~+8.0}_{~-5.7}$   & $udsc\bar{c}$ &~$\Xi_b^- \to J/\psi \Lambda K^-$ \\
$P_{cs}(4338)^0$ \cite{LHCb:2022ogu}  &    $4338.2 \pm 0.7 \pm 0.4$   &  $7.0 \pm 1.2 \pm 1.3$    & $udsc\bar{c}$ & $B^-\to J/\psi \Lambda \bar{p}$\\
\toprule
\end{tabular}
\end{table}

Very recently, Belle Collaboration found evidence of the $P_{cs}(4459)^0$ state with a significance of 3.3 standard deviations,    including statistical and systematic uncertainties. They  measure the mass and width of the $P_{cs}(4459)^0$ to be $(4471.7 \pm 4.8 \pm 0.6)~$ MeV and $(21.9 \pm 13.1 \pm 2.7)~$ MeV, respectively \cite{Belle:2025pey}. 
Along with the aforementioned hidden-charm pentaquark states, searches for doubly and triply strange hidden-charm pentaquarks are currently underway, with the CMS Collaboration recently observing the decay $\Lambda_b^0\to J/\psi\Xi^-K^+$~\cite{CMS:2024vnm}. Nevertheless, insufficient yield and inadequate resolution prevented the observation of a clear spectrum in the $J/\psi\Xi^-$ invariant mass. These results are significant in elucidating the underlying strong interaction processes in the hadronic decays of beauty baryons and the potential mechanism for forming exotic states. The observations in question have generated considerable excitement, leading to the intensified theoretical study of these hadrons. The objective of these theoretical studies was to investigate the properties of these states in order to gain insight into their natures and substructures. Moreover, some of these studies concentrated on the potential for the emergence of additional states, intending to provide insights that could inform future experimental observations. A variety of approaches and structural assumptions were employed to conduct a comprehensive investigation of the observed states. Details of these studies can be found in reviews~\cite{Esposito:2014rxa, Esposito:2016noz, Olsen:2017bmm, Lebed:2016hpi, Nielsen:2009uh, Brambilla:2019esw, Agaev:2020zad, Chen:2016qju, Ali:2017jda, Guo:2017jvc, Liu:2019zoy, Yang:2020atz, Dong:2021juy, Dong:2021bvy, Chen:2022asf, Meng:2022ozq}.
 
Despite the extensive research conducted since the initial observation of these states, our knowledge of their exact nature, internal structure, and quantum numbers remains incomplete. It is thus evident that further investigation into their properties is required. The present study examines the magnetic moments of the $P_{c}(4457)$   and related hidden-charm pentaquark states with and without strangeness ($[d d][u c] \bar c$, $[u u][s c] \bar c$, $[dd ][s c] \bar c$, $[s s][u c] \bar c$ and $[s s][d c] \bar c$), employing a comprehensive analysis that encompasses both the compact pentaquark configuration and $J^P = \frac{3}{2}^-$ quantum numbers. It is widely acknowledged that magnetic moments represent physical parameters that are directly correlated with the inner structure of the state under investigation.  Accordingly, the aforementioned parameters facilitate the extraction of insights concerning the internal structure of the hadron and the low-energy domain of QCD. Furthermore, magnetic moments represent an effective tool for investigating the dynamics of quarks and gluons within a hadron. This is due to the fact that it represents the leading-order response of a bound state to an external magnetic field. Despite the priceless insights they can provide, there are few investigations of the magnetic moments of hidden-charm/bottom pentaquarks in the available literature~\cite{Wang:2016dzu, Ozdem:2018qeh, Ortiz-Pacheco:2018ccl, Xu:2020flp, Ozdem:2021ugy, Ozdem:2021btf, Li:2021ryu, Ozdem:2023htj, Wang:2023iox, Ozdem:2022kei, Gao:2021hmv, Guo:2023fih,  Wang:2022nqs, Wang:2022tib, Ozdem:2024jty, Li:2024wxr, Li:2024jlq,  Ozdem:2024yel, Ozdem:2024rqx, Mutuk:2024ltc, Mutuk:2024jxf, PhysRevD.111.074038, Ozdem:2022iqk}. While the short lifetime of the $P_{c}$ states currently presents a significant challenge for measuring the magnetic moment, the accumulation of more extensive data from future experiments may facilitate the achievement of this goal.   The $\Delta^+(1232)$ baryon has also a very short lifetime, however, its magnetic moment was achieved through $\gamma N $ $ \rightarrow $ $ \Delta $ $\rightarrow $ $ \Delta \gamma $ $ \rightarrow$ $ \pi N \gamma $ process~ \cite{Pascalutsa:2004je, Pascalutsa:2005vq, Pascalutsa:2007wb}. A comparable process, $\gamma^{(*)}N $ $ \rightarrow $ $P_{c} $ $\rightarrow P_{c} \gamma$ $ \rightarrow $ $ J/\psi N \gamma$, may be employed to derive the magnetic moment of the $P_{c}$ pentaquarks.  Moreover, the magnetic moments of baryons containing two charm quarks have been calculated using techniques from lattice QCD~\cite{Can:2013zpa, Can:2013tna}.  The possibility exists for these analyses to be extended soon to encompass exotic states.

We organize this paper in the following manner: In Sec.~\ref{formalism} we present the results of a QCD light-cone sum rule analysis within an external background field method, which are conducted to calculate the magnetic moment of the considered states, which we have labeled as $P_c$. Additionally, we present an analysis of the results obtained from these analyses.  Sec.~\ref{numerical} is dedicated to a comprehensive numerical analysis of the magnetic moments of the $P_c$ states. Finally, this work ends with the summary in Sec.~\ref{summary}.

\begin{widetext}
 
\section{Theoretical Formalism for the magnetic moment}\label{formalism}

The QCD light-cone sum rule represents a widely accepted and productive methodology for the determination of the measurable properties of hadrons, including the calculation of their form factors, the analysis of their strong and weak decay properties, and their investigation in the context of radiative decays~\cite{Chernyak:1990ag, Braun:1988qv, Balitsky:1989ry}. The fundamental premise of the methodology is the calculation of the correlation function, which constitutes a pivotal element of the methodology. This is accomplished through two distinct approaches: the QCD approach and the hadronic approach. At the hadronic level, hadronic parameters, including residues, masses, and form factors, are utilized. In contrast, at the QCD level, parameters associated with QCD, such as the quark condensate, gluon condensate, and particle distribution amplitudes, among others, are employed. Subsequently, the Borel transformation and continuum subtraction are employed. The procedures above yield sum rules for the physics parameter to be calculated.

Following a brief introduction to the method, we may now proceed with the analysis of the physical parameter in question using this approach. As previously stated, the initial step is to define the correlation function of interest.  The correlation function to be employed in the analysis of the magnetic moment of the $P_{c}$ states is provided by the following formula:
\begin{eqnarray} \label{edmn01}
\Pi_{\mu \nu}(p,q)&=&i\int d^4x e^{ip \cdot x} \langle0|T\left\{J_{\mu}(x)\bar{J}_{\nu}(0)\right\}|0\rangle _\gamma \, ,
\end{eqnarray}
where  the $J_{\mu}(x)$ stands for interpolating current of the $P_{c}$ states,  $\gamma$ is the external electromagnetic field, and $q$ is the momentum of the photon.  The relevant expression for the $J_{\mu}(x)$ is presented as follows \cite{Wang:2015ixb,Wang:2019got}:

\begin{eqnarray}
J_{\mu}(x)&=&\frac{\mathcal A }{\sqrt{3}} \Big\{ \big[ {q_1}^T_d(x) C \gamma_\mu {q_1}_e(x) \big] \big[ {q_2}^T_f(x) C \gamma_5 c_g(x)\big]  C  \bar{c}^{T}_{c}(x) + 2
\big[ {q_1}^T_d(x) C \gamma_\mu {q_2}_e(x) \big] \big[ {q_1}^T_f(x) C \gamma_5 c_g(x)\big]  C  \bar{c}^{T}_{c}(x) \Big\} \, , 
\end{eqnarray}
where $ \mathcal A = \varepsilon_{abc}\varepsilon_{ade} \varepsilon_{bfg}$ with   
$a$, $b$, $c$, $d$, $e$, $f$ and $g$ being color indices; and the $C$ is the charge conjugation operator. The quark content of the $P_c$ states is listed in Table \ref{quarkcon}.
\begin{table}[htp]
	\addtolength{\tabcolsep}{10pt}
		\begin{center}
		\caption{The quark content of the $P_c$ states.}
	\label{quarkcon}
\begin{tabular}{lccccccccc}
	   \hline\hline
	   \\
  States& $[u u][d c] \bar c$&$[d d][u c] \bar c$&$[u u][s c] \bar c$&$[dd ][s c] \bar c$&$[s s][u c] \bar c$&$[s s][d c] \bar c$\\
  \\
\hline\hline
$q_1$&  u & d& u&d&s&s\\
$q_2$&  d & u& s&s&u&d\\
	   \hline\hline
\end{tabular}
\end{center}
\end{table}

In this step of the analysis, we will demonstrate how to derive the magnetic moment calculation in the form of hadronic parameters. In order to achieve this, a complete set with the same quantum numbers as the interpolating currents in Eq. (\ref{edmn01}) is incorporated into the correlation function. The resulting outcome is as follows:

\begin{eqnarray}\label{edmn02}
\Pi^{Had}_{\mu\nu}(p,q)&=&\frac{\langle0\mid J_{\mu}(x)\mid
P_{c}(p_2,s)\rangle}{[p_2^{2}-m_{P_{c}}^{2}]}\langle P_{c}(p_2,s)\mid
P_{c}(p_1,s)\rangle_\gamma\frac{\langle P_{c}(p_1,s)\mid
\bar{J}_{\nu}(0)\mid 0\rangle}{[p_1^{2}-m_{P_{c}}^{2}]},
\end{eqnarray}
where $p_1 = p+q$, $p_2=p$. As can be observed from the provided formulas, matrix elements such as $\langle0\mid J_{\mu}(x)\mid P_{c}(p_2,s)\rangle$, $\langle {P_{c}}(p_1,s)\mid
\bar{J}_{\nu}^{P_{c}}(0)\mid 0\rangle$, and $\langle P_{c}(p_2,s)\mid P_{c}(p_1,s)\rangle_\gamma$  emerge and are necessary for the remainder of the analysis. The matrix elements $\langle0\mid J_{\mu}(x)\mid P_{c}(p_2,s)\rangle$ and  $\langle {P_{c}}(p_1,s)\mid
\bar{J}_{\nu}^{P_{c}}(0)\mid 0\rangle$ are presented in the form provided below: 
\begin{eqnarray}
\label{lambdabey}
\langle0\mid J_{\mu}(x)\mid P_{c}(p_2,s)\rangle &=&\lambda_{P_{c}}u_{\mu}(p_2,s),\\
\langle {P_{c}}(p_1,s)\mid
\bar{J}_{\nu}(0)\mid 0\rangle &=& \lambda_{{P_{c}}}\bar u_{\nu}(p_1,s),
\end{eqnarray}
where $\lambda_{P_{c}}$  is current coupling of the $P_{c}$ states; $u_{\mu}(p_2,s)$ and $\bar u_{\nu}(p_1,s)$ are the Rarita-Schwinger spinors, which describe spin-3/2 hadrons and satisfying the Dirac equation:
 $(\pslash-m)u_{\mu}=0 $, $\gamma_\mu u_{\mu}=0$, $p_\mu u_{\mu}=0$.

The explicit form of the remaining matrix element $\langle P_{c}(p_2,s)\mid P_{c}(p_1,s)\rangle_\gamma$ is given as follows \cite{Weber:1978dh,Nozawa:1990gt,Pascalutsa:2006up,Ramalho:2009vc}: 

\begin{eqnarray}
\langle P_{c}(p_2,s)\mid P_{c}(p_1,s)\rangle_\gamma &=&-e \,\bar
u_{\mu}(p_2,s)\left\{F_{1}(q^2)g_{\mu\nu}\eslash-
\frac{1}{2m_{P_{c}}}\left
[F_{2}(q^2)g_{\mu\nu} \eslash\qslash+F_{4}(q^2)\frac{q_{\mu}q_{\nu} \eslash\qslash}{(2m_{P_{c}})^2}\right]
\right.\nonumber\\&+&\left.
\frac{F_{3}(q^2)}{(2m_{P_{c}})^2}q_{\mu}q_{\nu}\eslash\right\} u_{\nu}(p_1,s),\label{matelpar}
\end{eqnarray}
where $\varepsilon$ is the photon's polarization vector, and $F_i (q^2)$ are transition form factors.  By employing the Eqs.~(\ref{edmn02}) to (\ref{matelpar}) and making the requisite simplifications, the expressions for the magnetic moment of the $P_c$ states in conjunction with hadronic parameters are as follows:
\begin{align}\label{fizson}
 \Pi^{Had}_{\mu\nu}(p,q)&=-\frac{\lambda_{{P_{c}}}^{2}}{[p_1^{2}-m_{_{P_{c}}}^{2}][p_2^{2}-m_{_{P_{c}}}^{2}]}
 \big(\pslash_1+m_{P_{c}}\big)
 \bigg[g_{\mu\nu}
-\frac{1}{3}\gamma_{\mu}\gamma_{\nu}-\frac{2\,p_{1\mu}p_{1\nu}}
{3\,m^{2}_{P_{c}}}+\frac{p_{1\mu}\gamma_{\nu}-p_{1\nu}\gamma_{\mu}}{3\,m_{P_{c}}}\bigg] \bigg\{F_{1}(q^2)g_{\mu\nu}\eslash  \nonumber\\
& -
\frac{1}{2m_{P_{c}}}
\Big[F_{2}(q^2)g_{\mu\nu}\eslash\qslash +F_{4}(q^2) \frac{q_{\mu}q_{\nu}\eslash\qslash}{(2m_{P_{c}})^2}\Big]+\frac{F_{3}(q^2)}{(2m_{P_{c}})^2}
 q_{\mu}q_{\nu}\eslash\bigg\}
 \big(\pslash_2+m_{P_{c}}\big)
 \bigg[g_{\mu\nu}-\frac{1}{3}\gamma_{\mu}\gamma_{\nu}-\frac{2\,p_{2\mu}p_{2\nu}}
{3\,m^{2}_{P_{c}}}\nonumber\\
&+\frac{p_{2\mu}\gamma_{\nu}-p_{2\nu}\gamma_{\mu}}{3\,m_{P_{c}}}\bigg].
\end{align}
Here, a summation over the Rarita-Schwinger spinors is performed as follows: 
\begin{align}\label{raritabela}
\sum_{s}u_{\mu}(p,s)\bar u_{\nu}(p,s)=-\big(\pslash+m_{P_{c}}\big)\Big[g_{\mu\nu}
-\frac{1}{3}\gamma_{\mu}\gamma_{\nu}-\frac{2\,p_{\mu}p_{\nu}}
{3\,m^{2}_{P_{c}}}+\frac{p_{\mu}\gamma_{\nu}-p_{\nu}\gamma_{\mu}}{3\,m_{P_{c}}}\Big].
\end{align}

In general, the expression of the desired physical parameters can be achieved through the use of hadronic quantities under the previously established Eq. (\ref{fizson}). However, at this juncture, an additional step can be taken to further enhance the reliability and consistency of the analytical procedure. 
%
%
The interpolating current $J_\mu$ couples not only to spin-3/2 hadron states but also to spin-1/2 states. In other words $J_\mu$ has nonzero overlap with spin-1/2 states~\cite{Belyaev:1982cd, Belyaev:1993ss}. Therefore, in the expressions of Eq. (\ref{fizson}), both spin-1/2 hadrons contribute, and not all Lorentz structures are independent. In order to define the matrix element of the vacuum and the interpolating current between the spin-1/2 hadrons, it is possible to proceed as follows: 
\begin{equation}\label{spin12}
\langle 0 \mid J_{\mu}(0)\mid H(p,s=1/2)\rangle=(A  p_{\mu}+B\gamma_{\mu})u(p,s=1/2), 
\end{equation}
where $H(p,s=1/2)$ denotes the spin-$1/2$ hadron states. 

As demonstrated by the equation, the undesired effects associated with spin-1/2 hadrons have been found to be proportional to both $\gamma_\mu$ and $p_\mu$.   In order to remove the undesired contamination from the spin-1/2 hadrons and to obtain a correlation function comprising solely independent structures, we have devised the following ordering of the Dirac matrices:  $\gamma_{\mu}\pslash\eslash\qslash\gamma_{\nu}$.  Subsequently, in order to guarantee the exclusion of these irrelevant elements from the analysis, it is necessary to eliminate any terms with $\gamma_\mu$ at the beginning and $\gamma_\nu$ at the end, or that are directly proportional to $p_{2\mu}$ or $p_{1\nu}$~\cite{Belyaev:1982cd}. The outcome of the examination conducted at the hadron level, after the implementation of the aforementioned procedures, is presented in the following form:
\begin{eqnarray}
\label{final phenpart}
\Pi^{Had}_{\mu\nu}(p,q)&=&\frac{\lambda_{_{P_{c}}}^{2}}{[(p+q)^{2}-m_{_{P_{c}}}^{2}][p^{2}-m_{_{P_{c}}}^{2}]}
\bigg[  g_{\mu\nu}\pslash\eslash\qslash \,F_{1}(q^2) 
-m_{P_{c}}g_{\mu\nu}\eslash\qslash\,F_{2}(q^2)+
\frac{F_{3}(q^2)}{2m_{P_{c}}}q_{\mu}q_{\nu}\eslash\qslash\, \nonumber\\&+&
\frac{F_{4}(q^2)}{4m_{P_{c}}^3}(\varepsilon.p)q_{\mu}q_{\nu}\pslash\qslash \,+
\mathrm{other~independent~structures} \bigg].
\end{eqnarray}

It is more appropriate to express the form factors in Eq.~(\ref{final phenpart}) in terms of the magnetic form factor, \( G_{M}(q^2) \), as it represents an experimentally measurable quantity. The corresponding expression is presented below.
~\cite{Weber:1978dh,Nozawa:1990gt,Pascalutsa:2006up,Ramalho:2009vc}: 

\begin{eqnarray}
G_{M}(q^2) &=& \left[ F_1(q^2) + F_2(q^2)\right] ( 1+ \frac{4}{5}
\tau ) -\frac{2}{5} \left[ F_3(q^2)  +
F_4(q^2)\right] \tau \left( 1 + \tau \right), 
\end{eqnarray}
  where $\tau
= -\frac{q^2}{4m^2_{P_{c}}}$.  At $q^2=0$, the $G_M(0)$ regarding the $F_i(0)$ form factors is given by

\begin{eqnarray}\label{mqo1}
G_{M}(0)&=&F_{1}(0)+F_{2}(0).
\end{eqnarray}
 
In light of the central role of magnetic moments in our analysis, it becomes crucial to formulate the magnetic moment based on the form factors introduced earlier.  
The magnetic moment, denoted by ($\mu_{P_c}$), is extracted from the previously discussed term using the procedure outlined below:

 \begin{eqnarray}\label{mqo2}
\mu_{P_{c}}&=&\frac{e}{2m_{P_{c}}}G_{M}(0).
\end{eqnarray}

The calculation of the analysis in connection with the hadronic parameters has now been completed. The next phase of the calculations, conducted at the quark-gluon level, can now commence. To perform this phase of the analysis, the relevant interpolating currents are injected into the correlation function and all the relevant contractions are performed with the help of Wick's theorem. The result of this procedure is as follows:
\begin{align}
\label{QCD1}
\Pi^{QCD}_{\mu\nu}(p,q)&= \frac{i}{3}\,\mathcal {A} \mathcal{A^\prime}
\int d^4x e^{ip\cdot x}\langle 0|
\Big\{
 \nonumber\\
&
- \mbox{Tr}\Big[  \gamma_\mu S_{q_1}^{ee^\prime}(x) \gamma_\nu   C S_{q_1}^{dd^\prime \mathrm{T}}(x) C\Big]
 \mbox{Tr}\Big[ \gamma_5 S_c^{gg^\prime}(x) \gamma_5 C  S_{q_2}^{ff^\prime \mathrm{T}}(x)C \Big] 
 \nonumber\\
&
+ \mbox{Tr}\Big[  \gamma_\mu S_{q_1}^{ed^\prime}(x) \gamma_\nu C  S_{q_1}^{de^\prime \mathrm{T}}(x) C\Big]
 \mbox{Tr}\Big[ \gamma_5 S_c^{gg^\prime}(x) \gamma_5 C  S_{q_2}^{ff^\prime \mathrm{T}}(x)C  \Big] 
\nonumber\\
&
 +2 \mbox{Tr} \Big[ \gamma_5 S_c^{gg^\prime}(x) 
\gamma_5 C S_{q_1}^{ef^\prime \mathrm{T}}(x)  C \gamma_\mu S_{q_1}^{dd^\prime}(x) \gamma_\nu C  S_{q_2}^{fe^\prime \mathrm{T}}(x) C\Big]
\nonumber\\
&
 -2 \mbox{Tr} \Big[ \gamma_5 S_c^{gg^\prime}(x) 
\gamma_5 C S_{q_1}^{df^\prime \mathrm{T}}(x)  C \gamma_\mu S_{q_1}^{ed^\prime}(x) \gamma_\nu C  S_{q_2}^{fe^\prime \mathrm{T}}(x) C\Big]
\nonumber\\
&
-4 \mbox{Tr}\Big[  \gamma_\mu S_{q_2}^{ee^\prime}(x) \gamma_\nu C  S_{q_1}^{dd^\prime \mathrm{T}}(x) C\Big]
 \mbox{Tr}\Big[ \gamma_5 S_c^{gg^\prime}(x) \gamma_5 C  S_{q_1}^{ff^\prime \mathrm{T}}(x)C \Big] 
 \nonumber\\
&
+4 \mbox{Tr}\Big[  \gamma_\mu S_{q_2}^{ee^\prime}(x) \gamma_\nu C  S_{q_1}^{fd^\prime \mathrm{T}}(x) C\Big]
 \mbox{Tr}\Big[ \gamma_5 S_c^{gg^\prime}(x) \gamma_5 C  S_{q_1}^{df^\prime \mathrm{T}}(x)C  \Big] 
\nonumber\\
&
+2 \mbox{Tr} \Big[ \gamma_5 S_c^{gg^\prime}(x) 
\gamma_5 C S_{q_2}^{ef^\prime \mathrm{T}}(x)  C \gamma_\mu S_{q_1}^{dd^\prime}(x) \gamma_\nu C  S_{q_1}^{fe^\prime \mathrm{T}}(x) C\Big]
\nonumber\\
&
 -2 \mbox{Tr} \Big[ \gamma_5 S_c^{gg^\prime}(x) 
\gamma_5 C S_{q_2}^{ef^\prime \mathrm{T}}(x)  C \gamma_\mu S_{q_1}^{de^\prime}(x) \gamma_\nu C  S_{q_1}^{fd^\prime \mathrm{T}}(x) C\Big]
\Big \} 
\Big(C S_c^{c^{\prime}c \mathrm{T}} (-x) C \Big)
|0 \rangle_\gamma .
\end{align}

 The $S_{q}(x)$ and $S_{c}(x)$ in Eq. (\ref{QCD1}) are the related full quark  propagators, which are given  as~\cite{Yang:1993bp, Belyaev:1985wza}
\begin{align}
\label{edmn13}
S_{q}(x)&= S_q^{free}(x) - \frac{\langle \bar qq \rangle }{12} \Big(1-i\frac{m_{q} \xslash}{4}   \Big)- \frac{ \langle \bar qq \rangle }{192}
m_0^2 x^2  \Big(1-i\frac{m_{q} \xslash}{6}   \Big)
-\frac {i g_s }{16 \pi^2 x^2} \int_0^1 du \, G^{\mu \nu} (ux)
\bigg[\bar u \rlap/{x} 
\sigma_{\mu \nu} + u \sigma_{\mu \nu} \rlap/{x}
 \bigg],\\
\nonumber\\
%
S_{Q}(x)&=S_Q^{free}(x)
-i\frac{m_{Q}\,g_{s} }{16\pi ^{2}}  \int_0^1 du \,G^{\mu \nu}(ux)\bigg[ (\sigma _{\mu \nu }{\xslash}
+{\xslash}\sigma _{\mu \nu }) 
    \frac{K_{1}\big( m_{Q}\sqrt{-x^{2}}\big) }{\sqrt{-x^{2}}}
 +2\sigma_{\mu \nu }K_{0}\big( m_{Q}\sqrt{-x^{2}}\big)\bigg],
 \label{edmn14}
\end{align}%
with  
\begin{align}
 S_q^{free}(x)&=\frac{1}{2 \pi x^2}\Big(i \frac{\xslash}{x^2}- \frac{m_q}{2}\Big),\\
 \nonumber\\
 S_c^{free}(x)&=\frac{m_{c}^{2}}{4 \pi^{2}} \bigg[ \frac{K_{1}\big(m_{c}\sqrt{-x^{2}}\big) }{\sqrt{-x^{2}}}
+i\frac{{\xslash}~K_{2}\big( m_{c}\sqrt{-x^{2}}\big)}
{(\sqrt{-x^{2}})^{2}}\bigg], 
\end{align}
where \( G_{\mu\nu} \) represents the gluon field strength tensor, with \( G_{\mu\nu}^{ab} = G_A^{\mu\nu} t_A^{ab} \), \( t_A = \frac{\lambda_A}{2} \), and \( G^2 = G_A^{\mu\nu} G_{A\mu\nu} \). The indices \( A \) range from 1 to 8, corresponding to the Gell-Mann matrices, and \( K_n \) are the modified Bessel functions of the second kind. Here, we use the following integral representation of the modified Bessel function of the second kind:     
\begin{equation}\label{b2}
K_n(m_Q\sqrt{-x^2})=\frac{\Gamma(n+ 1/2)~2^n}{m_Q^n \,\sqrt{\pi}}\int_0^\infty dt~\cos(m_Qt)\frac{(\sqrt{-x^2})^n}{(t^2-x^2)^{n+1/2}}.
\end{equation}

The photon interacts with quarks in two distinct ways: at short distances, which is referred to as the perturbative contribution, and at long distances, which is known as the non-perturbative contribution. The perturbative contributions pertain to the short-distance interaction of the photon with all quark fields. In contrast, the non-perturbative contributions are concerned with the long-distance interaction of the photon with light quark fields.

  To account for perturbative contributions in the computations, one of the light or heavy quark propagators that are in interaction with the photon must be modified according to the following substitution:
  
\begin{align}
\label{free}
S^{free}(x) \rightarrow \int d^4z\, S^{free} (x-z)\,\rlap/{\!A}(z)\, S^{free} (z)\,.
\end{align}
Here we use $ A_\mu(z)=-\frac{1}{2}\, F_{\mu\nu}(z)\, z^\nu $ where  the electromagnetic field strength tensor is written as $ F_{\mu\nu}(z)=-i(\varepsilon_\mu q_\nu-\varepsilon_\nu q_\mu)\,e^{iq.z} $. 
After applying the manipulations described above, the expressions for \( S_q^{\text{free}}(x) \) and \( S_c^{\text{free}}(x) \) are obtained in the following forms:

\begin{eqnarray}\label{sfreepert}
&& S_q^{free}(x)=\frac{e_q}{32 \pi^2 x^2}\Big(\varepsilon_\alpha q_\beta-\varepsilon_\beta q_\alpha\Big)
 \Big(\xslash\sigma_{\alpha \beta}+\sigma_{\alpha\beta}\xslash\Big),\\
&& S_c^{free}(x)=-i\frac{e_c m_c}{32 \pi^2}
\Big(\varepsilon_\alpha q_\beta-\varepsilon_\beta q_\alpha\Big)
\Big[2\sigma_{\alpha\beta}K_{0}\Big( m_{c}\sqrt{-x^{2}}\Big)
 +\frac{K_{1}\Big( m_{c}\sqrt{-x^{2}}\Big) }{\sqrt{-x^{2}}}
 \Big(\xslash\sigma_{\alpha \beta}+\sigma_{\alpha\beta}\xslash\Big)\Big].
\end{eqnarray}

As an example, let us briefly illustrate how these transformations are applied by considering the first trace expression in Eq.~(\ref{QCD1}):
\begin{align}
\label{QCD11}
\Pi^{QCD-Pert}_{\mu\nu}(p,q)&= \frac{i}{3}\,
\mathcal {A} \mathcal{A^\prime}
\int d^4x e^{ip\cdot x}\langle 0|
\Big\{
 \nonumber\\
&
- \mbox{Tr}\Big[  \gamma_\mu S_{q_1}^{free}(x) \gamma_\nu  C S_{q_1}^{dd^\prime \mathrm{T}}(x) C\Big]
 \mbox{Tr}\Big[ \gamma_5 S_c^{gg^\prime}(x) \gamma_5 C  S_{q_2}^{ff^\prime \mathrm{T}}(x)C \Big]
 \nonumber\\
&
- \mbox{Tr}\Big[  \gamma_\mu S_{q_1}^{ee^\prime}(x) \gamma_\nu  C S_{q_1}^{free^\mathrm{T}}(x) C\Big]
 \mbox{Tr}\Big[ \gamma_5 S_c^{gg^\prime}(x) \gamma_5 C  S_{q_2}^{ff^\prime \mathrm{T}}(x)C \Big]
 \nonumber\\
&
- \mbox{Tr}\Big[  \gamma_\mu S_{q_1}^{ee^\prime}(x) \gamma_\nu  C S_{q_1}^{dd^\prime \mathrm{T}}(x) C\Big]
 \mbox{Tr}\Big[ \gamma_5 S_c^{free}(x) \gamma_5 C  S_{q_2}^{ff^\prime \mathrm{T}}(x)C \Big]
 \nonumber\\
&
- \mbox{Tr}\Big[  \gamma_\mu S_{q_1}^{ee^\prime}(x) \gamma_\nu  C S_{q_1}^{dd^\prime \mathrm{T}}(x) C\Big]
 \mbox{Tr}\Big[ \gamma_5 S_c^{gg^\prime}(x) \gamma_5 C  S_{q_2}^{free^\mathrm{T}}(x)C \Big]
 \nonumber\\
&
- \mbox{Tr}\Big[  \gamma_\mu S_{q_1}^{ee^\prime}(x) \gamma_\nu  C S_{q_1}^{dd^\prime \mathrm{T}}(x) C\Big]
 \mbox{Tr}\Big[ \gamma_5 S_c^{gg^\prime}(x) \gamma_5 C  S_{q_2}^{ff^\prime \mathrm{T}}(x)C \Big]
 \Big \} 
\Big(C S_c^{free^\mathrm{T}} (-x) C \Big) 
\nonumber\\
&
 + \mbox{Other Traces} \,\,
|0 \rangle_\gamma .
\end{align}

It should be noted that all possible contributions are taken into account in the above equations. In the first line of Eq.~(\ref{QCD11}), the photon couples perturbatively to the light quark propagator, while the remaining four propagators are treated as full propagators. Similarly, in the third line of Eq.~(\ref{QCD11}), the photon interacts perturbatively with the heavy quark propagators, with the other propagators considered as full, and so on.

To include non-perturbative elements in the computations, it is essential to modify one of the light quark propagators that interact with the photon at a long-distance, in adherence to the following replacement:
\begin{align}
\label{edmn15}
S_{\alpha\beta}^{ab}(x) \rightarrow -\frac{1}{4} \big[\bar{q}^a(x) \Gamma_i q^b(0)\big]\big(\Gamma_i\big)_{\alpha\beta},
\end{align}
where   $\Gamma_i = \mathrm{1}, \gamma_5, \gamma_\mu, i\gamma_5 \gamma_\mu, \sigma_{\mu\nu}/2$.   
In this scenario, the correlation function is given by:
\begin{align}
\label{QCD22}
\Pi^{QCD-Nonpert}_{\mu\nu}(p,q)&= \frac{i}{3}\,\mathcal {A} \mathcal{A^\prime}
\int d^4x e^{ip\cdot x}\langle 0|
\Big\{
 \nonumber\\
&
- \mbox{Tr}\Big[  \gamma_\mu \Gamma_i \gamma_\nu   CS_{q_1}^{dd^\prime \mathrm{T}}(x) C\Big]
 \mbox{Tr}\Big[ \gamma_5 S_c^{gg^\prime}(x) \gamma_5 C  S_{q_2}^{ff^\prime \mathrm{T}}(x)C \Big] 
 \Big(-
\frac{1}{4}\Big)\Big( \bar q ^e (x) \Gamma_i  q^{e^{\prime }}(0)\Big)
\Big(C S_c^{c^{\prime}c \mathrm{T}} (-x) C \Big)
\nonumber\\
&
- \mbox{Tr}\Big[  \gamma_\mu S_{q_1}^{ee^\prime}(x) \gamma_\nu   C \Gamma_i^{ \mathrm{T}} C\Big]
 \mbox{Tr}\Big[ \gamma_5 S_c^{gg^\prime}(x) \gamma_5 C  S_{q_2}^{ff^\prime \mathrm{T}}(x)C \Big]
 \Big(-
\frac{1}{4}\Big)\Big( \bar q ^d (x) \Gamma_i  q^{d^{\prime }}(0)\Big)
\Big(C S_c^{c^{\prime}c \mathrm{T}} (-x) C \Big)
 \nonumber\\
&
- \mbox{Tr}\Big[  \gamma_\mu S_{q_1}^{ee^\prime}(x) \gamma_\nu   C S_{q_1}^{dd^\prime \mathrm{T}}(x) C\Big]
 \mbox{Tr}\Big[ \gamma_5 S_c^{gg^\prime}(x) \gamma_5 C  \Gamma_i^{ \mathrm{T}}C \Big] \Big(-
\frac{1}{4}\Big)
 \Big( \bar q ^f (x) \Gamma_i  q^{f^{\prime }}(0)\Big)
\Big(C S_c^{c^{\prime}c \mathrm{T}} (-x) C \Big)
 \Big \}
 \nonumber\\
&
 + \mbox{Other Traces} \,\,
|0 \rangle_\gamma.
\end{align}

By substituting the light quark propagators and employing the relation \( \bar{q}^a(x) \Gamma_i q^{a'}(0) \rightarrow \frac{1}{3} \delta^{aa'} \bar{q}(x) \Gamma_i q(0) \), Eq.~(\ref{QCD22}) is rewritten as follows:
\begin{align}
\label{QCD23}
\Pi^{QCD-Nonpert}_{\mu\nu}(p,q)&= -\frac{i}{3}\,\mathcal {A} \mathcal{A^\prime}
\int d^4x e^{ip\cdot x}\langle 0|
\Big\{
 \nonumber\\
&
- \mbox{Tr}\Big[  \gamma_\mu \Gamma_i \gamma_\nu   C S_{q_1}^{dd^\prime \mathrm{T}}(x) C\Big]
 \mbox{Tr}\Big[ \gamma_5 S_c^{gg^\prime}(x) \gamma_5 C  S_{q_2}^{ff^\prime \mathrm{T}}(x)C \Big] 
\Big(C S_c^{c^{\prime}c \mathrm{T}} (-x) C \Big)
\nonumber\\
&
- \mbox{Tr}\Big[  \gamma_\mu S_{q_1}^{ee^\prime}(x) \gamma_\nu   C \Gamma_i^{ \mathrm{T}} C\Big]
 \mbox{Tr}\Big[ \gamma_5 S_c^{gg^\prime}(x) \gamma_5 C  S_{q_2}^{ff^\prime \mathrm{T}}(x)C \Big]
\Big(C S_c^{c^{\prime}c \mathrm{T}} (-x) C \Big)
 \nonumber\\
&
- \mbox{Tr}\Big[  \gamma_\mu S_{q_1}^{ee^\prime}(x) \gamma_\nu  C S_{q_1}^{dd^\prime \mathrm{T}}(x) C\Big]
 \mbox{Tr}\Big[ \gamma_5 S_c^{gg^\prime}(x) \gamma_5 C  \Gamma_i^{ \mathrm{T}}C \Big]
\Big(C S_c^{c^{\prime}c \mathrm{T}} (-x) C \Big)
 \Big \} 
 \nonumber\\
& \times \frac{1}{12} \langle \gamma(q) |\bar q(x)\Gamma_i q(0)|0\rangle
 + \mbox{Other Traces}.
\end{align}

In addition, when a light quark couples non-perturbatively to a photon, it is also possible for a gluon to be emitted from one of the other four quark propagators.
The resulting expression after carrying out these calculations is given below:
\begin{align}
\label{QCD33}
\Pi^{QCD-Nonpert}_{\mu\nu}(p,q)&= -\frac{i}{3}\,\mathcal {A} \mathcal{A^\prime}
\int d^4x e^{ip\cdot x}\langle 0|
\Big\{
 \nonumber\\
&
- \mbox{Tr}\Big[  \gamma_\mu \Gamma_i \gamma_\nu   C S_{q_1}^{dd^\prime \mathrm{T}}(x) C\Big]
 \mbox{Tr}\Big[ \gamma_5 S_c^{gg^\prime}(x) \gamma_5 C  S_{q_2}^{ff^\prime \mathrm{T}}(x)C \Big] 
\Big(C S_c^{c^{\prime}c \mathrm{T}} (-x) C \Big)
\Big[\Big(\delta^{ed}\delta^{e^\prime d^\prime}-\frac{1}{3}\delta^{ee^\prime}\delta^{dd^\prime}\Big)
\nonumber\\
& 
+
\Big(\delta^{eg}\delta^{e^\prime g^\prime}-\frac{1}{3}\delta^{ee^\prime}\delta^{gg^\prime}\Big)
+
\Big(\delta^{ef}\delta^{e^\prime f^\prime}-\frac{1}{3}\delta^{ee^\prime}\delta^{ff^\prime}\Big)
+
\Big(\delta^{ec}\delta^{e^\prime c^\prime}-\frac{1}{3}\delta^{ee^\prime}\delta^{cc^\prime}\Big)
\Big]
\nonumber\\
&
- \mbox{Tr}\Big[  \gamma_\mu S_{q_1}^{ee^\prime}(x) \gamma_\nu   C \Gamma_i^{ \mathrm{T}} C\Big]
 \mbox{Tr}\Big[ \gamma_5 S_c^{gg^\prime}(x) \gamma_5 C  S_{q_2}^{ff^\prime \mathrm{T}}(x)C \Big]
\Big(C S_c^{c^{\prime}c \mathrm{T}} (-x) C \Big) 
\Big[\Big(\delta^{de}\delta^{d^\prime e^\prime}-\frac{1}{3}\delta^{dd^\prime}\delta^{ee^\prime}\Big)
\nonumber\\
& 
+
\Big(\delta^{dg}\delta^{d^\prime g^\prime}-\frac{1}{3}\delta^{dd^\prime}\delta^{gg^\prime}\Big)
+
\Big(\delta^{df}\delta^{d^\prime f^\prime}-\frac{1}{3}\delta^{dd^\prime}\delta^{ff^\prime}\Big)
+
\Big(\delta^{dc}\delta^{d^\prime c^\prime}-\frac{1}{3}\delta^{dd^\prime}\delta^{cc^\prime}\Big)
\Big]
 \nonumber\\
&
- \mbox{Tr}\Big[  \gamma_\mu S_{q_1}^{ee^\prime}(x) \gamma_\nu C  S_{q_1}^{dd^\prime \mathrm{T}}(x) C\Big]
 \mbox{Tr}\Big[ \gamma_5 S_c^{gg^\prime}(x) \gamma_5 C  \Gamma_i^{ \mathrm{T}}C \Big]
\Big(C S_c^{c^{\prime}c \mathrm{T}} (-x) C \Big) 
\Big[\Big(\delta^{df}\delta^{d^\prime f^\prime}-\frac{1}{3}\delta^{ee^\prime}\delta^{dd^\prime}\Big)
\nonumber\\
& 
+
\Big(\delta^{fe}\delta^{f^\prime e^\prime}-\frac{1}{3}\delta^{ff^\prime}\delta^{ee^\prime}\Big)
+
\Big(\delta^{fg}\delta^{f^\prime g^\prime}-\frac{1}{3}\delta^{ff^\prime}\delta^{gg^\prime}\Big)
+
\Big(\delta^{fc}\delta^{f^\prime c^\prime}-\frac{1}{3}\delta^{ff^\prime}\delta^{cc^\prime}\Big)
\Big]
 \Big \} 
 \nonumber\\
& \times \frac{1}{32} \langle \gamma(q) |\bar q(x)\Gamma_i G_{\mu\nu}(x) q(0)|0\rangle
 + \mbox{Other Traces},
\end{align}
where we inserted  
\begin{align}
\label{QCDES5}
 \bar q^a(x)\Gamma_i G_{\mu\nu}^{bb'}(x) q^{a'}(0)\rightarrow \frac{1}{8}\Big(\delta^{ab}\delta^{a'b'}
 -\frac{1}{3}\delta^{aa'}\delta^{bb'}\Big)\bar q(x)\Gamma_i G_{\mu\nu}(x) q(0).
\end{align}

 In the context of the non-perturbative analysis, the terms $\langle \gamma(q)\vel \bar{q}(x) \Gamma_i G_{\mu\nu}(x)q(0) \ver 0\rangle$  and $\langle \gamma(q)\vel \bar{q}(x) \Gamma_i q(0) \ver 0\rangle$ appear and are pivotal to the remainder of the computations. These terms are expressed about the distribution amplitudes of the photon, along with associated parameters, as detailed in Ref.~\cite{Ball:2002ps}. These transformations ensure that all possible contributions are included in the analysis, while also allowing for the individual quark contributions to be examined, as all the relevant transformations are proportional to the quark charges (\( e_{q} \) and \( e_Q \)). 
Since these aspects of the analysis are technical and have been standardized, we have not included further detail here. Those interested in this topic may wish to consult the Refs.~\cite{Ozdem:2022vip, Ozdem:2022eds}, which provides more detailed information and a more comprehensive account of the procedures in question. Eqs.~(\ref{free}) and (\ref{QCD33}) have been employed to incorporate both perturbative and non-perturbative contributions into the analysis, following the established methodology.

As a result of the above-mentioned methodology, the subsequent equality is obtained:
\begin{align}
 \Pi^{QCD}_{\mu\nu}(p,q)=\Pi^{QCD-Pert}_{\mu\nu}(p,q)+\Pi^{QCD-Nonpert}_{\mu\nu}(p,q). 
\end{align}

After deriving the correlation functions at both the hadronic and quark-gluon levels, the next procedure involves establishing the sum rules for the magnetic moment. The analytical expressions corresponding to the magnetic moments of the \( P_c \) states are provided in the following equation:
\begin{align}
 &\mu_{P_c}\, \lambda^2_{P_c} = e^{\frac{m^2_{P_{c}}}{\mathrm{M^2}}}\, \rho (\mathrm{M^2},\mathrm{s_0}),
\end{align}
where 
\begin{align}
  \rho (\mathrm{M^2},\mathrm{s_0}) &=  F_1(\mathrm{M^2},\mathrm{s_0}) -\frac{1}{m_{P_{cs}}} F_2(\mathrm{M^2},\mathrm{s_0}),
 \end{align}
with
\begin{align}
\label{F1son}
 F_1 (\mathrm{M^2},\mathrm{s_0})&= -\frac{e_c}{2^{25}\times 3^3 \times 5^3 \times 7^2  \pi^7} \Big[3752 m_{q_1} m_c I[0, 6] - 855 I[0, 7]\Big]
 \nonumber\\
 &+\frac{C_1 C_2 C_3}{2^{17}\times 3^6 \times 5  \pi^3} \Big[45 e_c m_{q_1} m_c I[0, 1] - 11 (e_c + e_{q_1}) I[0, 2]\Big]
 \nonumber\\
 &+\frac{C_1 C_2^2}{2^{17}\times 3^5   \pi^3} \Big[ (e_c - 2 e_{q_2}) m_{q_1} m_c I[0, 1]\Big]
 \nonumber\\
  &+\frac{C_1 C_2}{2^{22}\times 3^7 \times 5  \pi^5}
 \Bigg [m_ 0^2 \Big (-220 (e_ {q_ 1} + e_ {q_ 2}) m_ {q_ 1} - 
      3 (197 e_ {q_ 1} + 251 e_ {q_ 2}) m_c + 
      e_c (671 m_ {q_ 1} + 402 m_c)\Big) I[0, 
     2] 
     \nonumber\\
 & + \Big (-901 e_c m_ {q_ 1} + 60 e_ {q_ 1} m_ {q_ 1} + 
      832 e_ {q_ 2} m_ {q_ 1} - 934 e_c m_c - 24 e_ {q_ 1} m_c + 
      624 e_ {q_ 2} m_c\Big) I[0, 3]\Bigg]
 \nonumber\\
%
 &+\frac{C_1 C_3}{2^{21}\times 3^7 \times 5   \pi^5}\Bigg [-3 m_ 0^2 \Big (33 (e_c + e_ {q_ 1}) m_ {q_ 1} + 34 e_c m_c + 
      22 e_ {q_ 1} m_c\Big) I[0, 2] - 
   2 \Big (271 e_c m_ {q_ 1} - 268 e_ {q_ 1} m_ {q_ 1} \nonumber\\
 &+ 26 e_c m_c + 
      12 e_ {q_ 1} m_c\Big) I[0, 3]\Bigg]
 \nonumber\\
 &+\frac{ C_2^2\, e_c}{2^{15}\times 3^5 \times 5   \pi^3}
 \Big [ 55 m_0^2 m_{q_1} m_c I[0, 2] + 12 m_{q_1} m_c I[0, 3] + 9 [0, 4]\Big] 
     \nonumber\\
 &+\frac{ C_1}{2^{26}\times 3^7 \times 5^2   \pi^7}
 \Big[ (15 (531 e_c + 175 e_{q_1} - 215 e_{q_2}) m_{q_1} m_c I[0, 
    4] + (57 e_c - 1316 e_{q_1} - 496 e_{q_2}) I[0, 5]) \Big]
 \nonumber\\
 &+\frac{e_c}{2^{21}\times 3^5 \times 5^2  \pi^5}\Bigg[15 m_ 0^2 \Big (9 m_ {q_ 1}(7 C_ 2 + 8 C_ 3)  - 
     19 m_c (2 C_ 2 + C_ 3) \Big) I[0, 4] + 
  4 \Big (-27 m_ {q_ 1} (C_ 2 + 7 C_ 3)  
  \nonumber\\
 &+ 
     28 m_c (2 C_ 2 + C_ 3) \Big) I[0, 5]\Bigg],\\
     \nonumber
\end{align} 
\begin{align}
 F_2 (\mathrm{M^2},\mathrm{s_0})&= \frac{e_c\, m_c}{2^{22}\times 3^5 \times  7^2  \pi^7} \Big[82880 m_{q_1} m_c I[0, 6] - 18063 I[0, 7]\Big]
 \nonumber\\
 &+\frac{C_1 C_2 C_3\, m_c}{2^{17}\times 3^6 \times 5   \pi^3} \Big[   
 15 e_c m_{q_1} m_c I[0, 1] - 7 e_c I[0, 2] + 7 e_{q_1} I[0, 2] + 
 4 (2 e_c - 5 e_{q_1}) I[1, 1]
 \Big]
 \nonumber\\
 &+\frac{C_1 C_2^2\, m_c}{2^{16}\times 3^6  \pi^3} \Big[  18 e_c m_ {q_ 1} m_c I[0, 1] - 36 e_ {q_ 2} m_ {q_ 1} m_c I[0, 1] - 
 e_c I[0, 2] + 10 e_ {q_ 2} I[0, 2] + 2 (e_c - 10 e_ {q_ 2}) I[1, 1]  \Big]
 \nonumber\\
 &+\frac{C_1 C_2\, m_c}{2^{21}\times 3^5 \times 5   \pi^5}
 \Bigg [  -5 e_ {q_ 1} \Big (4 (8 m_ {q_ 2} - 11 m_c) I[0, 3] + 
    15 m_ 0^2 \big ((8 m_ {q_ 2} - 3 m_c) I[0, 2] + 
       4 (-2 m_ {q_ 2} + 3 m_c) 
       \nonumber\\
 & \times I[1, 1]\big) + 
    12 (11 m_ {q_ 2} - 20 m_c) I[1, 2]\Big) + 
 e_c \Big ((-1259 m_ {q_ 2} + 640 m_c) I[0, 3] + 
    15 m_ 0^2 \big ((65 m_ {q_ 2} - 84 m_c) I[0, 2] \nonumber\\
 &- 
       8 (m_ {q_ 2} - 18 m_c) I[1, 1]\big) + 
    360 (m_ {q_ 2} - 5 m_c) I[1, 2]\Big) + 
 e_ {q_ 2} \Big (4 (114 m_ {q_ 2} + 145 m_c) I[0, 3] + 
    15 m_ 0^2 \big ((20 m_ {q_ 2} \nonumber\\
 &+ 51 m_c) I[0, 2] - 
       8 (10 m_ {q_ 2} + 21 m_c) I[1, 1]\big) + 
    24 (59 m_ {q_ 2} + 95 m_c) I[1, 2]\Big)  \Bigg]
 \nonumber\\
 &+\frac{C_1 C_3\, m_c}{2^{20}\times 3^7    \pi^5}\Bigg [  e_ {q_ 1} \Big (8 (8 m_ {q_ 1} - m_c) I[0, 3] + 
    9 m_ 0^2 \big ((7 m_ {q_ 1} + 4 m_c) I[0, 2] - 
       4 (5 m_ {q_ 1} + m_c) I[1, 1]\big) + 
    48 (5 m_ {q_ 1} \nonumber\\
%
 & + m_c) I[1, 2]\Big) - 
 e_c \Big (59 (2 m_ {q_ 1} - m_c) I[0, 3] + 
    9 m_ 0^2 \big ((7 m_ {q_ 1} + 13 m_c) I[0, 2] - 
       4 (2 m_ {q_ 1} + 3 m_c) I[1, 1]\big) + 
    6 (16 m_ {q_ 1} \nonumber\\
 &+ 15 m_c) I[1, 2]\Big)\Bigg]
 \nonumber\\
 &-\frac{ C_2^2\, e_c\,m_c}{2^{14}\times 3^5 \times 5   \pi^3}
 \Big [  75 m_0^2 m_{q_1} m_c I[0, 2] + 20 m_{q_1} m_c I[0, 3] + 9 I[0, 4]\Big] 
     \nonumber\\
&+\frac{ C_1\, m_c}{2^{28}\times 3^6 \times 5^2   \pi^7}
 \Bigg[ -160 (53 e_ {q_ 1} + 107 e_ {q_ 2}) m_ {q_ 1} m_c I[0, 4] + 
 6537 e_c I[0, 5] - 8 (77 e_ {q_ 1} + 52 e_ {q_ 2}) I[0, 5] 
 \nonumber\\
&- 
 1280 (20 e_ {q_ 1} + 29 e_ {q_ 2}) m_ {q_ 1} m_c I[1, 3] + 
 320 e_c m_ {q_ 1} m_c (-31 I[0, 4] + 84 I[1, 3]) + 
 180 (9 e_c + 29 e_ {q_ 1} + 4 e_ {q_ 2}) I[1, 4] \Bigg]
 \nonumber\\
 &+\frac{e_c\, m_c}{2^{20}\times 3^4 \times 5^2   \pi^5}\Big[  
 120 m_0^2 \Big(-3 m_{q_1}(C_2 + C_3)  + m_c(2 C_2 + C_3) \Big) I[0, 
   4] + \Big(9 m_{q_1}(5 C_2 + 32 C_3)  - 46 m_c(2 C_2 \nonumber\\
   &+ C_3)\Big) I[0, 5]\Big],
   \label{F2son}
\end{align}

 \noindent  where $C_1 =\langle g_s^2 G^2\rangle$ is gluon condensate;  $C_2 =\langle \bar q_1 q_1 \rangle$ and $C_3 =\langle \bar q_2 q_2 \rangle$ are corresponding light-quark condensates. The expressions provided above include only the terms that make a significant contribution to the numerical values of the magnetic moments. To maintain clarity, terms not presented here are also considered in the numerical computations. 
 The function $\mathrm{I}[n,m]$ is given as 
\begin{align}
 \mathrm{I}[n,m]&= \int_{\mathcal M}^{\mathrm{s_0}} ds~ e^{-s/\mathrm{M^2}}~
 s^n\,(s-\mathcal M)^m,
 \end{align}
 \end{widetext} 
 \noindent where $\mathcal{M}= 4m_c^2$ for the $P_c(4457) $ and $[d d][u c] \bar c$ states;  $\mathcal{M}= (2m_c+m_s)^2$ for the $[u u][s c] \bar c$ and $[dd ][s c] \bar c$ states; and,   $\mathcal{M}= (2m_c+2m_s)^2$ for the $[s s][u c] \bar c$ and $[s s][d c] \bar c$ states.

The Borel transformations are executed according to the well-established formulas listed below: 
\begin{align}
 \mathcal{B}\bigg\{ \frac{1}{\big[ [p^2-m^2_i][(p+q)^2-m_f^2] \big]}\bigg\} \rightarrow e^{-m_i^2/M_1^2-m_f^2/M_2^2}
\end{align}
at the hadron level, 
\begin{align}
 \mathcal{B}\bigg\{ \frac{1}{\big(m^2- \bar u p^2-u(p+q)^2\big)^{\alpha}}\bigg\} \rightarrow (M^2)^{(2-\alpha)} \delta (u-u_0)e^{-m^2/M^2},
\end{align}
 at the quark-gluon level, the following expressions are employed:
\begin{align*}
 {M^2}= \frac{M_1^2 M_2^2}{M_1^2+M_2^2}, ~~~
 u_0= \frac{M_1^2}{M_1^2+M_2^2}.
\end{align*}
In this framework, \( M_1^2 \) and \( M_2^2 \) represent the Borel parameters corresponding to the initial and final pentaquark states, respectively. Since the same pentaquark state is present in both the initial and final channels, it is appropriate to choose \( M_1^2 = M_2^2 = 2M^2 \) and \( u_0 = 1/2 \). This choice provides a symmetric treatment of both sides of the correlation function and is sufficient to effectively suppress the contributions from higher excited states and the continuum.

\section{Numerical illustrations}\label{numerical}

 This section presents the results of a numerical analysis of the QCD light-cone sum rule, which was conducted in order to make predictions regarding the magnetic moments of the $P_{c}$ states.  A prerequisite to undertaking the numerical analysis of the QCD light-cone sum rule is to ascertain the numerical values of many parameters. In order to facilitate a comprehensive analysis, a set of values has been adopted for the parameters in question:  
$m_s =93.4^{+8.6}_{-3.4}\,\mbox{MeV}$, $m_c = 1.27 \pm 0.02\,\mbox{GeV}$~\cite{ParticleDataGroup:2022pth},  $m_{P_{c}(4457)} = 4457.3 \pm 0.6^{+4.1}_{-1.7}$ MeV~\cite{Aaij:2019vzc},  
$m_{[dd][uc]\bar c} = 4.47 \pm 0.11$ GeV~\cite{Wang:2019got}, 
$m_{[uu][sc]\bar c} = 4.51 \pm 0.12$ GeV~\cite{Wang:2015ixb}, 
$m_{[dd][sc]\bar c} = 4.51 \pm 0.12$ GeV~\cite{Wang:2015ixb},
$m_{[ss][uc]\bar c} = 4.60 \pm 0.11$ GeV~\cite{Wang:2015ixb}, 
$m_{[ss][dc]\bar c} = 4.60 \pm 0.11$ GeV~\cite{Wang:2015ixb}, 
$\lambda_{P_{c}(4457)}= (2.41 \pm 0.38) \times 10^{-3} \mbox{GeV}^6$ \cite{Wang:2019got},  
$\lambda_{[dd][uc]\bar c}= (2.41 \pm 0.38) \times 10^{-3} \mbox{GeV}^6$~\cite{Wang:2019got},  
$\lambda_{[uu][sc]\bar c}=(2.75 \pm 0.45) \times 10^{-3} \mbox{GeV}^6$ \cite{Wang:2015ixb}, 
$ \lambda_{[dd][sc]\bar c}=(2.75 \pm 0.45) \times 10^{-3} \mbox{GeV}^6$ \cite{Wang:2015ixb},
$\lambda_{[ss][uc]\bar c}=(3.19 \pm 0.50) \times 10^{-3} \mbox{GeV}^6$ \cite{Wang:2015ixb}, 
$ \lambda_{[ss][dc]\bar c}=(3.19 \pm 0.50) \times 10^{-3} \mbox{GeV}^6$ \cite{Wang:2015ixb}, 
$\langle \bar uu\rangle = 
\langle \bar dd\rangle=(-0.24 \pm 0.01)^3\,\mbox{GeV}^3$, $\langle \bar ss\rangle = (0.8 \pm 0.1)\, \langle \bar uu\rangle$ $\,\mbox{GeV}^3$ \cite{Ioffe:2005ym},
$m_0^{2} = 0.8 \pm 0.1 \,\mbox{GeV}^2$ \cite{Ioffe:2005ym},  and 
$\langle g_s^2G^2\rangle = 0.48 \pm 0.14~ \mbox{GeV}^4$~\cite{Narison:2018nbv}. 
In numerical calculations, we fix $m_u$ =$m_d$ = 0 and $m_s^2=0$, but taking into account terms proportional to $m_s$.   
To perform further computations, it is essential to utilize the photon DAs and their explicit form, together with the necessary numerical values, as outlined in Ref.~\cite{Ball:2002ps}.

\begin{figure}[htp]
\centering
\subfloat[]{\includegraphics[width=0.46\textwidth]{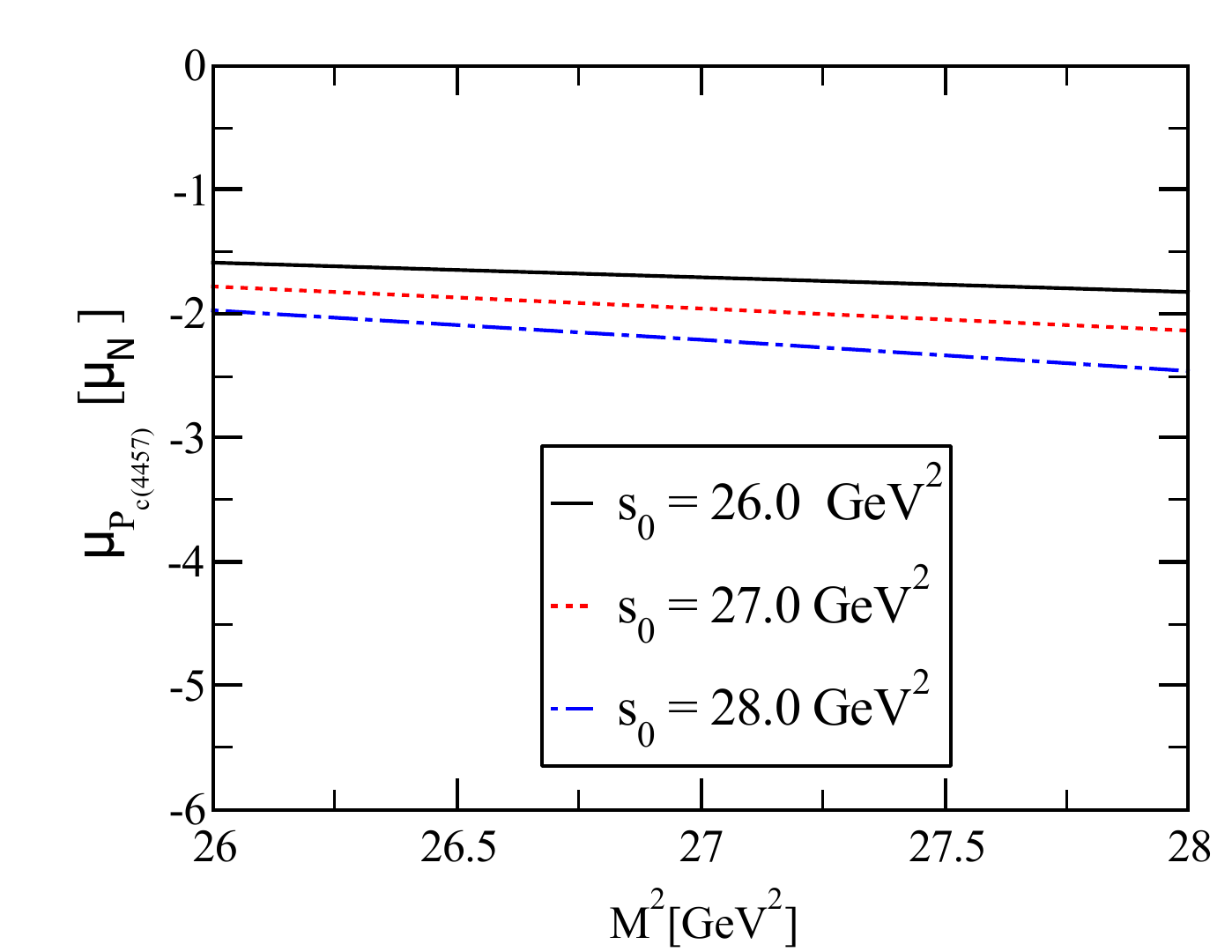}} ~~~~~~
\subfloat[]{\includegraphics[width=0.46\textwidth]{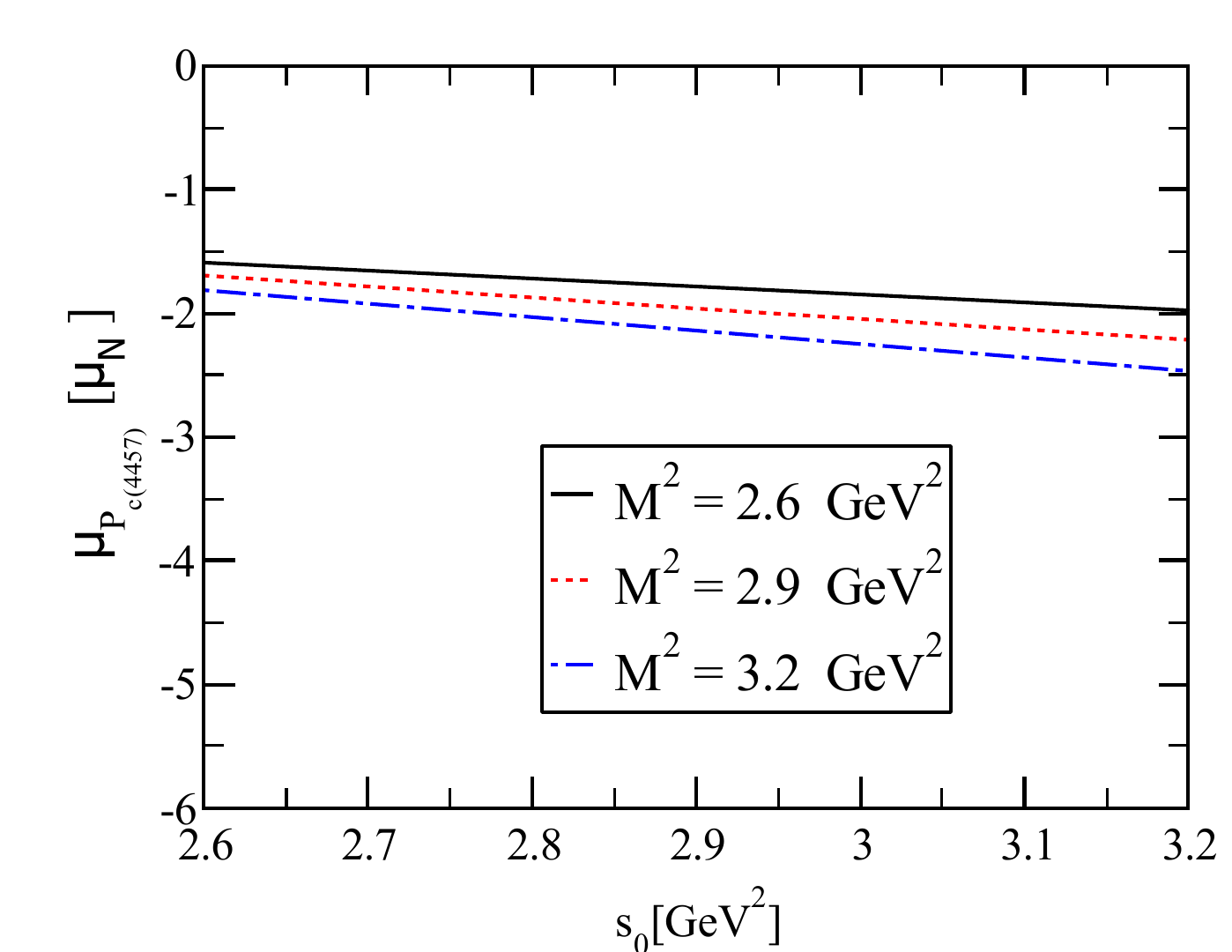}}\\
 \caption{The magnetic moments of the $P_{c}(4457)$ state versus $\mathrm{M^2}$ (left panel) and $\mathrm{s_0}$ (right panel).}
 \label{Msqfig}
  \end{figure}

In consideration of the previously outlined numerical input variables, two additional parameters are requisite for the implementation of our numerical analysis. These are the continuum threshold parameter, designated as $\mathrm{s_0}$, and the Borel parameter, represented by $\mathrm{M^2}$. In an ideal context, the numerical analysis would be conducted in a manner that is independent of the aforementioned parameters. Nevertheless, this is not a viable approach in practice. 
It is therefore necessary to establish a region of analysis in which the influence of parameter variation on the numerical results is to be regarded as insignificant. The range of applicability of these parameters is contingent upon the methodology employed, and the variation in numerical results concerning these specified parameters is minimal within the specified interval.  The working region for these parameters, which is the interval where the variation in our numerical results for these parameters is small, is subject to the constraints imposed by the methodology used. Such limitations are commonly designated as pole contribution (PC) and convergence of operator product expansion (CVG).  
 The aforementioned constraints are defined following the relevant formulas as follows:
\begin{align}
 \mathrm{PC} &=\frac{\rho (\mathrm{M^2},\mathrm{s_0})}{\rho (\mathrm{M^2},\infty)},~~~~~~\\
 \mathrm{CVG}&=\frac{\rho^{\mathrm{DimN}} (\mathrm{M^2},\mathrm{s_0})}{\rho (\mathrm{M^2},\mathrm{s_0})},
 \end{align}
 where $\rho_i^{\mbox{DimN}} (\rm{M^2},\rm{s_0})$ represent the highest dimensional terms in the $\rho_i (\rm{M^2},\rm{s_0})$. 
 As demonstrated in Eqs. (\ref{F1son})-(\ref{F2son}), our analysis includes combinations of condensates with various structures, such as $\langle g_s^2G^2\rangle  \langle \bar q_1 q_1\rangle \langle \bar q_2 q_2\rangle$ (Dim10), 
 $\langle g_s^2G^2\rangle  \langle \bar q_1 q_1\rangle^2$ (Dim10), 
 $m_0^2$ $\langle g_s^2G^2\rangle  \langle \bar q_1 q_1\rangle$ (Dim9),
 $m_0^2$ $\langle g_s^2G^2\rangle  \langle \bar q_2 q_2 \rangle$ (Dim9),
 $m_0^2$ $ \langle \bar q_1 q_1\rangle^2 $ (Dim8),
 $\langle g_s^2G^2\rangle  \langle \bar q_1 q_1\rangle$ (Dim7), 
 $\langle g_s^2G^2\rangle  \langle \bar q_2 q_2\rangle$ (Dim7), 
 $ \langle \bar q_1 q_1\rangle^2$ (Dim6),  
 $m_0^2$ $ \langle \bar q_1 q_1\rangle $ (Dim5), 
 $m_0^2$ $ \langle \bar q_2 q_2\rangle $ (Dim5), 
 $\langle g_s^2G^2\rangle$ (Dim4), $\langle \bar q_1 q_1\rangle$ (Dim3) and $\langle \bar q_2 q_2 \rangle$ (Dim3). As evident from the discussion, the highest-dimensional contributions in our analysis arise from dimension-8, dimension-9, and dimension-10 operators.   Accordingly, in the CVG analysis, the total dimension is taken as $\text{DimN} = \text{Dim}(8+9+10)$, and the analysis is performed accordingly.  
 Following the sum rules analysis, the CVG is required to be sufficiently small to ensure the convergence of the operator product expansion, whereas the PC is obliged to be sufficiently large in order to optimize the efficiency of the single-pole approach. The PC and CVG values obtained from the computational analysis are displayed in Table~\ref{parameter}, along with the designated working intervals of the $\mathrm{s_0}$, and $\mathrm{M^2}$ for the states under study. To guarantee the dependability of the derived working intervals,  we have taken the state of $P_c(4457)$ as an example, the variations of the calculated magnetic moment values about the specified auxiliary variables are plotted in Fig.~\ref{Msqfig}. As anticipated, the figure illustrates a slight discrepancy in the outcomes observed within these specified intervals.  
 Though the magnetic moments of these states show a slight dependence on these quantities, they remain within the permissible limits of this methodology, representing the primary source of uncertainty.

 %
  \begin{widetext}

  \begin{table}[htb!]
	\addtolength{\tabcolsep}{10pt}
	\caption{Numerical results of the magnetic moment of the $P_{c}$ states together with working intervals of helping parameters.}
	\label{parameter}
		\begin{center}
\begin{tabular}{l|ccccc}
	   \hline\hline
	   \\
   State& $\mu~[\mu_N]$& $\mathrm{s_0}$~\mbox{[GeV$^2$]}&$\mathrm{M^2}$~\mbox{[GeV$^2$]}&\mbox{CVG} $[\%]$&\mbox{PC} $[\%]$\\
   \\
\hline\hline
\\
$P_c(4457)$& $ -1.96^{+0.50}_{-0.37}$ &$ [26.0, 28.0]$         &  $ [2.6, 3.2] $& < 2&[35.1, 57.0]\\
\\
$[dd] [uc] \bar c$ & $ -2.04^{+0.46}_{-0.41}$ &$ [26.0, 28.0]$        &  $ [2.6, 3.2]$& < 2&[34.7, 56.9]\\
\\
$[uu] [sc] \bar c$& $ -2.08^{+0.53}_{-0.39}$ &$ [27.0, 29.0]$         &  $ [2.8, 3.4] $& < 2&[33.2, 53.8]\\
\\
$[dd] [sc] \bar c$& $ -2.13^{+0.53}_{-0.40}$ &$ [27.0, 29.0]$         &  $ [2.8, 3.4] $& < 2&[33.5, 54.4]\\
\\
$[ss] [uc] \bar c$& $ -2.29^{+0.53}_{-0.39}$ &$ [28.0, 30.0]$         &  $ [2.9, 3.5] $& < 2 &[35.0, 55.2]\\
\\
$[ss] [dc] \bar c$& $ -2.33^{+0.53}_{-0.41}$ &$ [28.0, 30.0]$         &  $ [2.9, 3.5] $& < 2&[35.1, 55.3]\\
\\
	   \hline\hline
\end{tabular}
\end{center}
\end{table}

\end{widetext}

 All requisite parameters for the numerical analysis have been identified and defined. The complete numerical results, inclusive of all inherent variabilities associated with the input parameters, are presented in Table~\ref{parameter}.  The findings that are obtained from the numerical results can be interpreted as follows:
 
 \begin{itemize}
 
 \item Magnetic moment size can be used to provide insight into the experimental accessibility of such entities. The magnetic moments determined for the hidden-charm pentaquarks are considerably large. The magnitude of these results points to the possibility that they may be achievable in future experiments.
 
 \item To undertake a deeper examination of the magnetic moment, the contribution of light quarks and the c-quark is also examined.   This can be accomplished by manipulating the respective charge factors (\( e_{q} \) and \( e_c \)) within the sum rules, which were intentionally preserved for this purpose. %
 For instance, the light-quark contribution to the magnetic moment can be isolated by setting $e_c = 0$ in Eqs.~(\ref{F1son})-(\ref{F2son}), leaving only terms proportional to $e_q$. This procedure is analogous to the method described in~\cite{Lee:2011nq}.  The results of this analysis are listed in Table \ref{parameter2}. It should be noted that the central values of all input parameters have been utilized to obtain this table. 
 From these results, the contributions of light quarks to the magnetic moment are found to be nearly negligible, accounting for only $\sim 10\%$ of the total value ($|\mu_q/\mu_{\text{total}}| \approx 0.1$). 
 The results obtained for the magnetic moment are dominated by the c-quark. 
  
\item  The contribution of charm- and light quarks to the magnetic moment have been observed to exhibit an inverse relationship. The signs of the magnetic moments demonstrate the interaction of the spin degrees of freedom of the quarks. The opposing signs of the charm- and light quarks magnetic moments indicate that their spins are anti-aligned in the $P_{c}$ states.
  
\item The $U$-symmetry violation in the predictions is acquired as a maximum of $11\%$. From these results, one can see that a reasonable $U$-symmetry violation is observed.
 
\item To obtain further insight, it would be beneficial to conduct a comparative analysis between the numerical values obtained and the existing literature on the subject. In Ref.~\cite{Ozdem:2021ugy}, the magnetic moments of the $P_{c}(4457)$ state were investigated within the framework of the QCD light-cone sum rules for both molecular and compact pentaquark configurations with $J^P = \frac{1}{2}^-$ quantum numbers. The resulting values are designated as $\mu_{P_{c}(4457)} =  2.78^{+0.94}_{-0.83}~\mu_N$ and $\mu_{P_{c}(4457)} =  0.88^{+0.32}_{-0.29}~\mu_N$ for the molecular and compact pentaquark configurations, respectively. In Ref.~\cite{Li:2021ryu}, the magnetic moment of the $P_{c}(4457)$ state was investigated in the quark model with the quantum numbers $J^P = \frac{3}{2}^-$, with and without coupled channel and $D$-wave effects. It has been suggested that these pentaquarks can be described within the molecular picture.   The resulting value is $\mu_{P_{c}(4457)} = (1.145-1.365)~\mu_N$.  The results exhibit considerable discrepancies, not only in magnitude but also in sign. The numerical results obtained in this study, when considered alongside existing literature, indicate that the magnetic moments of hidden-charm pentaquark states may offer insights into their underlying structures, which in turn can inform the distinction between their spin-parity quantum numbers. To get a more conclusive picture of these results, further studies are encouraged.

\item In general, it is expected that modifying the basis (e.g., charge, spin, and isospin) would not significantly affect the resulting data. However, this expectation may not be valid in the case of magnetic moments. 
The reason for this is that the electromagnetic characteristics of hadrons are directly tied to their internal structure. 
In terms of electromagnetic characteristics, altering the basis of the associated hadron results in a transformation of its internal structure, which can substantially impact the calculated results.
In Refs.~\cite{Wang:2016dzu, Gao:2021hmv, Ozdem:2024txt, Ozdem:2024dbq, Ozdem:2024jty, Azizi:2023gzv, Ozdem:2024rqx,  PhysRevD.111.074038, Ozdem:2022iqk}, the electromagnetic properties of multiquark states were examined under a range of assumptions, which revealed considerable deviations in the magnetic moments based on different internal configurations. Therefore, the choice of interpolating currents, and wave functions—or equivalently, the isospin, spin, and charge basis—can have a substantial impact on the magnetic moments of the studied hadrons.

\end{itemize}

  \begin{table}[htb!]
	\addtolength{\tabcolsep}{10pt}
	\caption{The contribution of light and heavy quarks to the magnetic moment of the $P_{c}$ states.}
	\label{parameter2}
		\begin{center}
\begin{tabular}{l|ccccc}
	   \hline\hline
	   \\
   Pentaquarks& $\mu_{q~[\mu_N]}$&  $\mu_{c}~[\mu_N]$& $\mu_{total}~[\mu_N]$\\
   \\
\hline\hline
\\
$P_{c}(4457)$& $ 0.23 $    &  $-2.19 $& $-1.96$\\
\\
$[dd] [uc] \bar c$& $ 0.21 $    &  $-2.24 $& $-2.04$\\
\\
$[uu] [sc] \bar c$& $ 0.22 $    &  $-2.30 $& $-2.08$\\
\\
$[dd] [sc] \bar c$ &$0.17 $   &  $ -2.30$&$ -2.13$\\
\\
$[ss] [uc] \bar c$& $0.21$      &  $-2.50$&$-2.29$\\
\\
$[ss] [dc] \bar c$& $0.20$    &  $ -2.53$&$-2.33$\\
\\
	   \hline\hline
\end{tabular}
\end{center}
\end{table}

\section{Summary}\label{summary}

We systematically study the electromagnetic properties of pentaquark states from different perspectives to better understand their nature, internal structure, and quantum numbers, determine their hadronization processes, and shed light on their true nature.  The present study examines the magnetic moments of the $P_{c}(4457)$ and related hidden-charm pentaquark states with and without strangeness ($[d d][u c] \bar c$, $[u u][s c] \bar c$, $[dd ][s c] \bar c$, $[s s][u c] \bar c$ and $[s s][d c] \bar c$), employing a comprehensive analysis that encompasses both the compact pentaquark configuration and $J^P = \frac{3}{2}^-$ quantum numbers. The present study compares the results regarding the magnetic moment of the $P_{c}(4457)$ pentaquark state with those reported in the existing literature. The numerical results obtained in this study, when considered alongside existing literature, indicate that the magnetic moments of hidden-charm pentaquark states may offer insights into their underlying structures, which in turn can inform the distinction between their spin-parity quantum numbers. It seems that for the future experimental search of the family of hidden-charm pentaquark states, the study of the electromagnetic properties of the hidden-charm pentaquark states can provide valuable information. A thorough analysis of the light and heavy quark contributions to the magnetic moment has been conducted. The analysis yielded results that indicate the contributions of light quarks to the magnetic moment are nearly negligible. The results obtained for the magnetic moment are found to be dominated by the charm quark. We hope that these revelations will motivate our experimental colleagues to probe further into the family of hidden-charm pentaquark states and to explore the inner structure of $P_c(4457)$ in future studies.


\bibliographystyle{elsarticle-num}
\bibliography{Pc4457MM.bib}
\end{document}